\documentclass[%
reprint,
 amsmath,amssymb,
aps,
pra,
]{revtex4-1}

\usepackage{graphicx}
\usepackage{dcolumn}
\usepackage{bm}

\begin{document}

\preprint{APS/123-QED}

\title{MAICRM : A general model for rapid simulation of hot dense plasmas}%

\author{Xiaoying Han}%
\email{han\_xiaoying@iapcma.ac.cn}
\author{Lingxiao Li}%
\author{Zhensheng Dai}%
\author{Wudi Zheng}%
\author{Peijun Gu}%
\author{Zeqing Wu}%
\affiliation{Institute of Applied Physics and Computational Mathematics, Beijing 100088, China}%

\date{\today}

\begin{abstract}
We propose a general model, Multi-Average Ion Collisional-Radiative Model (MAICRM), to rapid simulate the ionization and population distributions of hot dense plasmas. In MAICRM, the orbital occupation numbers of ions at the same charge stage are averaged and determined by the excitation and de-excitation processes; the populations of the average ions are determined by the ionization and recombination processes with the fixed orbital average occupation numbers in each ion. The calculated mean ionizations and charge state distributions of MAICRM are in general agreement with the other theoretical and experimental results especially for the mid- and high-density plasmas. Since MAICRM considers more detailed transitions and ionization balances than the average atom model and is faster than DCA/SCA models, this model has the advantage to be combined into hydrodynamic simulations.
\end{abstract}

\pacs{Valid PACS appear here}
\keywords{Suggested keywords}
\maketitle

\section{Introduction}
The charge state distribution (CSD) of hot dense plasmas is of primary interest of a great deal of experimental and theoretical attention since they are the key values for the hydrodynamic simulations, spectroscopic diagnostics and laser plasma interaction (LPI) simulations in the applications such as inertial confinement fusion (ICF) \cite{Lindl1995} and z-pinch plasmas\cite{Whitney2001}. The CSD of plasma is governed by a balance between processes such as ionization, recombination, excitation and radiative decay, as well as two-body processes such as autoionization and dielectronic recombination. For the high density plasma, the plasma approaches local thermodynamic equilibrium (LTE) and the CSD and the occupation numbers of the orbitals satisfy the Saha-Boltzmann equations. In the low-density limit, coronal equilibrium (CE) can be assumed and the CSD is evaluated by the CE equations. For an arbitrary density plasma the CSD and the orbital occupations are determined by the solution of a set of coupled collisional-radiative (CR) rate equations involving excitation/de-excitation and photoionization/recombination atomic processes.

As demonstrated in the 9th NLTE code comparison workshop\cite{NLTE9-2017}, according to the coarseness of the statistical treatment of the atomic levels, all the models can be roughly classified into a few categories including average-atom (AA) codes, detailed configuration accounting (DCA) and/or Super-transition-arrays (STA) codes, detailed level accounting (DLA) codes and hybrid codes. In the detailed atomic models such as DCA\cite{UTA1988}, STA\cite{STA1989} and DLA\cite{Ralchenko2001}, it is crucial to consider a sufficiently large number of energy levels in each ionization stage; this leads to sometimes unmanageable thus various approximations, e.g., neglecting multi-excited or inner-shell excited states, are adopted. In addition to the completion, the computation time is another key point especially for the time dependent cases. In general the main computation time in the models spends on the rate and absorption/emission coefficient calculations, which is proportional to $(N_{CS}\times N_{C}\times N_{O})$. Here $N_{CS}$, $N_{C}$ and $N_{O}$ are the numbers of the considered charge states, configurations in each charge state and the orbital bases. Thus AAM is of big advantage since  $(N_{CS}\times N_{C})^{\textsc{AAM}}=1$ while in other detailed models $(N_{CS}\times N_{C})=(Z+1)\times N_{C}$ and $N_{C}$ may be equal to tens or hundreds if including the single, double and multi-excited states. So the radiative hydrodynamic codes usually choose AAM as the combined inline atomic model despite its coarseness treatment of the atomic levels.With the progress of the experiment and simulation abilities some more detailed models such as SCA\cite{SCRAM2007} and DCA\cite{DCA1990} are carried out in the hydrodynamic simulations. Ref.\cite{JONES2017} shows the differences of the simulated emissivity and capsule bang time by using different inline atomic methods. In their comparisons, many mixed factors such as the numbers and the detail degrees of the involved energy levels are attributed to the differences. The influences of different fixed factors are difficult to be distinguished.

Considering the ionization/recombination processes are usually slower than the excitation/de-excitation processes if the electron density is not too low, the two kinds of processes can be decoupled. Here we propose a general method, i.e., Multi-Average Ion Collisional-Rdiative Model (MAICRM), to simulate the hot dense plasmas by separately solving the occupation numbers of the orbitals and the fraction populations of the ionic charge states. More specifically, in MAICRM, all the configurations at the same ionic stage, namely having the same number of bound electrons, are averaged and represented by an average ion and for a single element plasma, only (Z+1) average ions are taken into account. In this model, the average occupation numbers of each average ion are determined by the excitation and de-excitation processes and the fraction populations of all the average ion are determined by the ionization and recombination processes with the fixed non-integral average occupation numbers. Because of $(N_{CS}\times N_{C})^{\textsc{MAICRM}}=(Z+1)\times 1$ the computation time of MAICRM is shorter than those of DCA/SCA calculations by $N_{C}$ times, which is usually equal to tens or hundreds. On the other hand, in MAICRM all the configurations at one ionic stage are described by an average ion but the detailed ionization balances between the different ionic stage are considered, which is more detailed than AAM in which all the configurations in all ionic stages are averaged and represented by an average atom. Thus because of its shorter time consuming and more detailed treatment of the atomic levels the proposed MAICRM has the advantage to be combined into the hydrodynamic simulations. This paper is organized as following:
Section II introduces the theoretical method and Section III shows the comparisons of the mean ionization and CSDs of Fe, Xe, Au plasmas between the results of MAICRM with the other theoretical and experimental results.

\section{Theoretical method}
\subsection{Average ion}
In the present MAICRM, 65 single orbital bases $nlj$ or $nl$ are chosen and listed in table \ref{tab:orbitalbases} where $n, l$ and $j$ are the principal quantum number, angular momentum and total angular momentum. For $n\leq5$, the orbital bases are relativistic and for $5<n\leq10$, the orbital bases are non-relativistic. The energy levels of the orbital bases and their transition matrix elements are calculated by RSCF (relativistic self-consistent-field) method\cite{TXM1995} and tabulated as an input database. For the non-relativistic orbitals the values are averaged from their related relativistic orbitals by their static weights $(2j+1)$. For each charge state, an average ion $\mathbf{\Lambda^{\theta}}$ is labeled as
\begin{eqnarray}
{\mathbf{\Lambda^{\theta}}\equiv (n_{1}l_{1}j_{1})^{\mathbf{\Omega}^{\theta}_{1}}(n_{2}l_{2}j_{2})^{\mathbf{\Omega}^{\theta}_{2}} \cdots (n_{i^{\theta}_{max}}l_{i^{\theta}_{max}})^{\mathbf{\Omega}^{\theta}_{i^{\theta}_{max}}},}
\end{eqnarray}
where $\theta$ is the number of bound electrons in the average ion and $i^{\theta}_{max}$ is the max number of the single orbital bases for average ion $\mathbf{\Lambda^{\theta}}$. $\mathbf{\Omega}^{\theta}_{i}$ is the electron average occupying number of the $i$th orbital in $\mathbf{\Lambda^{\theta}}$, which is non-integral and satisfies $ 0 \leq \mathbf{\Omega}^{\theta}_{i} \leq \mathbf{g}_{i} $; here $\mathbf{g}_{i}$ is the statistic weight of the $i$th orbital. We define a vector $\mathbf{\Xi^{\theta}}$ to present the average ion $\mathbf{\Lambda^{\theta}}$
\begin{eqnarray}
{\mathbf{\Xi^{\theta}}\equiv \{\mathbf{\Omega}^{\theta}_{1},\mathbf{\Omega}^{\theta}_{2},\ldots,\mathbf{\Omega}^{\theta}_{i^{\theta}_{max}}\}.}
\end{eqnarray}

\begin{table}
\caption{\label{tab:orbitalbases}The single orbital bases}
\begin{ruledtabular}
\begin{tabular}{ll}
$1s_{1/2}$ \\
$2s_{1/2}, 2p_{1/2}, 2p_{3/2}$\\
$3s_{1/2}, 3p_{1/2}, 3p_{3/2}, 3d_{3/2}, 3d_{5/2}$\\
$4s_{1/2}, 4p_{1/2}, 4p_{3/2}, 4d_{3/2}, 4d_{5/2}, 4f_{5/2}, 4f_{7/2}$\\
$5s_{1/2}, 5p_{1/2}, 5p_{3/2}, 5d_{3/2}, 5d_{5/2}, 5f_{5/2}, 5f_{7/2}, 5g_{7/2}, 5g_{9/2}$\\
$6s, 6p, 6d, 6f, 6g, 6h$\\
$7s, 7p, 7d, 7f, 7g, 7h, 7i$\\
$8s, 8p, 8d, 8f, 8g, 8h, 8i, 8j$\\
$9s, 9p, 9d, 9f, 9g, 9h, 9i, 9j, 9k$\\
$10s, 10p, 10d, 10f, 10g, 10h, 10i, 10j, 10k$\\
\end{tabular}
\end{ruledtabular}
\end{table}

\subsection{Rate coefficients}
In our model, the rate coefficients of ten kinds of atomic processes, including photon excitation, electron collisional excitation, photon ionization, electron collisional ionization, autoionization and their reverse processes, use the same calculation formulas in Ref.\cite{Florido2009} with the integer occupation numbers of orbitals in each configuration replaced by the average occupation number $\mathbf{\Omega}^{\theta}_{i}$ in each average ion $\mathbf{\Lambda^{\theta}}$.

\subsection{Plasma effect}
In the dense plasma, screening effects due to neighboring electrons and ions modify the energy levels. The effect on the ionization potentials of bound states and level populations leads to the phenomenon of pressure ionization. In despite of the importance of pressure ionization to calculate the ionic abundances and level populations, most CR models take into account plasma effects in an approximate way via an effective lowering of the ionization potential or continuum lowering (CL). In our model the ionization potential $\mathbf{I}_{\theta}$ is lowered a quantity $\Delta \mathbf{I}_{\theta}$ to be $\mathbf{I}'_{\theta}=\mathbf{I}_{\theta}-\Delta \mathbf{I}_{\theta}$. Here we apply the formulation proposed by Stewart and Pyatt\cite{Stewart1966} and use the following formula
\begin{widetext}
\begin{eqnarray}
{\Delta \mathbf{I}_{\theta}=\frac{3}{2}\frac{I_{H}a_{0}}{\mathbf{r}^{\theta}}(\langle Z\rangle-\theta+1)\{[1+(\frac{\mathbf{D}}{\mathbf{r}^{\theta}})^{3}]^{\frac{2}{3}}-(\frac{\mathbf{D}}{\mathbf{r}^{\theta}})^{2}\},}\label{eq:eq18}
\end{eqnarray}
\end{widetext}
where $\mathbf{r^{\theta}}=[3(\langle Z\rangle-\theta+1)/(4\pi n_{e})]^{1/3}$ is the ion-sphere radius assuming the plasma composed of ions with $\theta$ electrons only, $\mathbf{D}=[4\pi (\langle Z\rangle+\langle Z^{2}\rangle)n_{ion}/T_{e}]^{-1/2}$ is the Debye length, $\langle Z\rangle$ is the average ionization of plasma, $\langle Z^{2}\rangle$ is the second-order moment of the population distribution. Once the lowered ionization potential $\mathbf{I}'_{\theta}$ are calculated for each charge state, the $i^{\theta}_{max}$ for each charge state are determined with the orbital energy $\mathbf{E^{\theta}_{\emph{i}}}<\mathbf{I}'_{\theta}$. Then a new set of $\mathbf{I}'_{\theta}$ is obtained based on the updated $i^{\theta}_{max}$. Thus $i^{\theta}_{max}$ and $\mathbf{I}'_{\theta}$ are calculated iteratively.
The CL correction may result in a significant reduction of the $i^{\theta}_{max}$ for some charge states if the density of the plasma is high.

\subsection{Average occupation number and charge state distribution}
In our model, the average occupying numbers of the orbitals and the populations of ionic charge states are calculated by two steps iteratively. Firstly, we calculate the average orbital occupation numbers $\mathbf{\Omega}^{\theta}_{i} (i=1,\ldots,i^{\theta}_{max})$ of every average ion $\mathbf{\Lambda}^{\theta}$ by solving a set of $i^{\theta}_{max}$ rate equations which are only relative to the excitation and de-excitation atomic processes in the ion. The coupled $i^{\theta}_{max}$ rate equations of $\mathbf{\Omega}^{\theta}_{i}$ for each average ion are
\begin{widetext}
\begin{eqnarray}
{\frac{d\mathbf{\Omega}^{\theta}_{i}}{d\mathbf{t}}=-\mathbf{\Omega}^{\theta}_{i}\sum_{j=1}^{i^{\theta}_{max}}(\mathbf{g}_{j}-\mathbf{\Omega}^{\theta}_{j})\mathbf{R^{E/D}_{\theta,\emph{i}\rightarrow \emph{j}}}+(\mathbf{g}_{i}-\mathbf{\Omega}^{\theta}_{i})\sum_{j=1}^{i^{\theta}_{max}}\mathbf{\Omega}^{\theta}_{j}\mathbf{R^{E/D}_{\theta,\emph{j}\rightarrow \emph{i}}},}\label{eq:rate1-1}
\end{eqnarray}
\end{widetext}
where $\mathbf{R^{E/D}_{\theta,\emph{i}\rightarrow \emph{j}}}$ are the excitation or de-excitation rate coefficients for the transition of $i$th orbital to $j$th orbital in the average ion $\mathbf{\Lambda}^{\theta}$. $\mathbf{\Omega}^{\theta}_{i}$ should fulfill the condition of charge conservation
\begin{eqnarray}
{\sum_{i=1}^{i^{\theta}_{max}}\mathbf{\Omega}^{\theta}_{i}=\theta~~(\theta = 0,\ldots,Z).}\label{eq:rate1-2}
\end{eqnarray}
Combining Eqs.~(\ref{eq:rate1-1}) and (\ref{eq:rate1-2}), a set of converged solutions $\{\Xi^{\theta},(\theta = 0,\ldots,Z)\}$ for all considered average ions are obtained.

Secondly, based on the set of fixed vectors $\{\Xi^{\theta},(\theta = 0,\ldots,Z)\}$, the coupled rate equations of the $(Z+1)$ average ions can be built as following
\begin{widetext}
\begin{eqnarray}
{\frac{d \mathbf{P_{\theta}}}{dt}=-\mathbf{P_{\theta}}\sum_{\tilde{\theta}=\theta\pm 1}\mathbf{R^{I/R}_{\theta\rightarrow \tilde{\theta}}}+\sum_{\tilde{\theta}=\theta\pm 1}\mathbf{P}_{\tilde{\theta}}\mathbf{R^{I/R}_{\tilde{\theta}\rightarrow \theta}} ~~~(\tilde{\theta}, \theta = 0,\ldots,Z) ,}\label{eq:rate2-1}
\end{eqnarray}
\end{widetext}
where $\mathbf{P}_{\theta}$ is the population of average ion $\mathbf{\Lambda^{\theta}}$. $\mathbf{R^{I/R}_{\theta\rightarrow \tilde{\theta}}}$ are the ionization or recombination rate coefficients of the average ion $\mathbf{\Lambda^{\theta}}$ to the average ion $\mathbf{\Lambda^{\tilde{\theta}}}$. $\mathbf{P}_{\theta}$ should fulfill the normalization condition
\begin{eqnarray}
{\sum_{\theta=0}^{Z}\mathbf{P}_{\theta}=1.}\label{eq:rate2-2}
\end{eqnarray}
Combining Eqs.~(\ref{eq:rate2-1}) and (\ref{eq:rate2-2}), a set of $\{ \mathbf{P}_{\theta} (\theta = 0, \ldots, Z) \}$ are obtained.
Then a new electron density value $n'_{e}=n_{i}\sum_{\theta=0}^{\mathbf{Z}}\mathbf{P}_{\theta}\cdot(Z-\theta)$ is calculated. With the new $n'_{e}$ the rate coefficients are updated and putted into Eq.~(\ref{eq:rate1-1}), then a new set of $\{\Xi'^{\theta},\theta = 0, \ldots, Z\}$ are obtained. With the updated $\{\Xi'^{\theta}\}$ a new set of $\{\mathbf{P}'_{\theta}\}$ are calculated by Eqs.~(\ref{eq:rate2-1}) and (\ref{eq:rate2-2}). The iterative procedure of updating $\{\Xi^{\theta}, {\mathbf{P}_{\theta}} (\theta = 0, \ldots, Z) \}$ is stopped when a converged $n_{e}$ and mean ionization $\langle Z\rangle$ reach.

\section{Results and discussions}
Using MAICRM the mean ion charge $\langle Z\rangle$ and CSD of mid- and high-Z plasmas in LTE and NLTE conditions are calculated and compared with other theoretical and experimental results.

\begin{figure}
\includegraphics[scale=0.35]{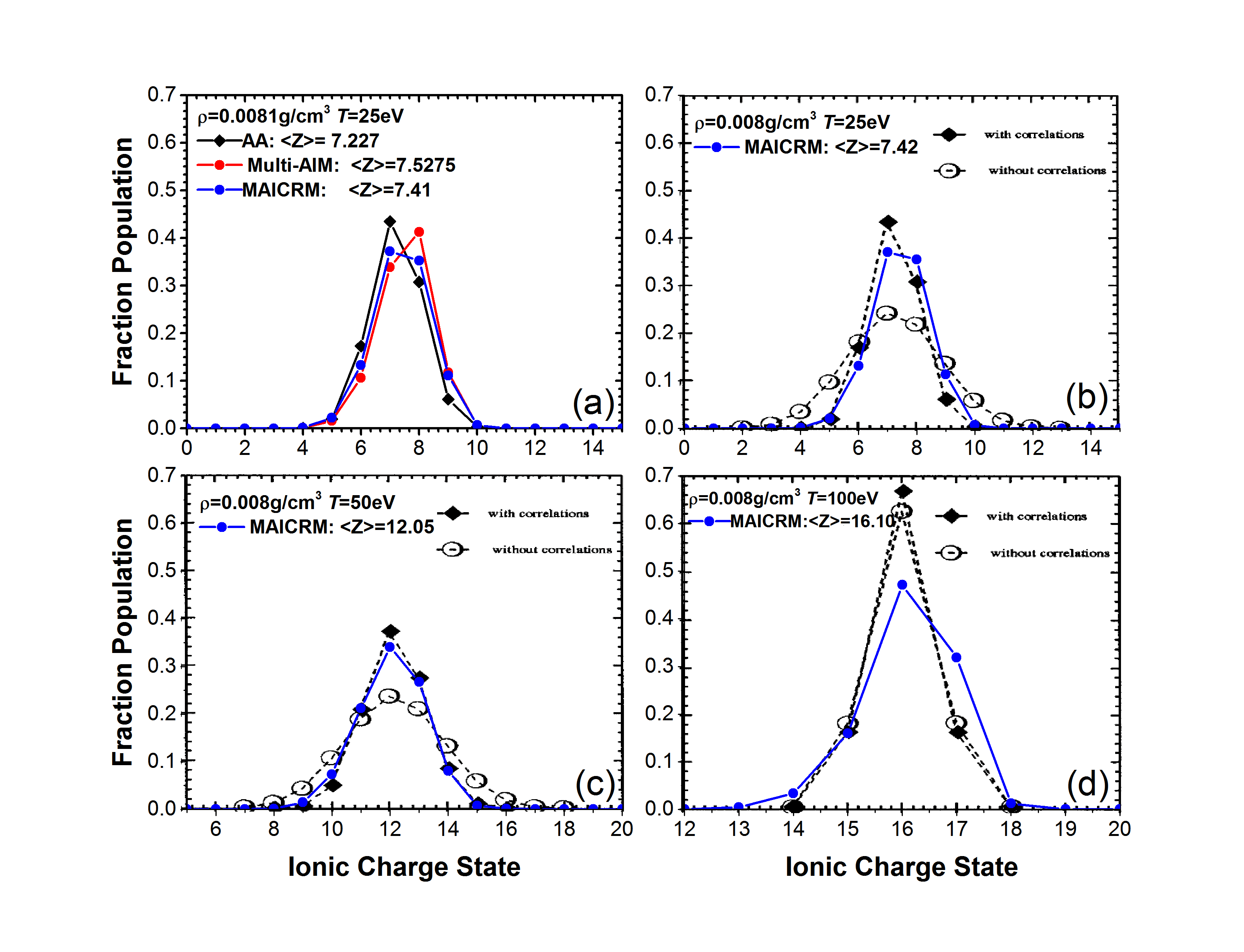}
\caption{\label{fig:FeLTE}(Color online) The comparison of the charge state distributions (CSDs) of Fe Plasma in LTE condition between MAICMR codes and the other calculations\cite{Faussurier1997,Kiyokawa2014}. }
\end{figure}

Fig.\ref{fig:FeLTE} shows the results of Fe plasma in LTE condition. In panel (a), for the plasma of the condition $\rho$=0.0081 g/cm$^{3}$, $T$=25eV, our calculated $\langle Z\rangle$ and CSD agree with the results of AAM\cite{Faussurier1997} and Multi-AIM\cite{Kiyokawa2014}. The other two models assume a non-integer orbital occupation determined by the Fermi distribution. The CSD of AAM is obtained from minor manipulations of the grand canonical partition function\cite{Faussurier1997}. The CSD of Multi-AIM is obtained self-consistently by means of minimizing the free energy of the whole system established by the finite temperature density functional theory\cite{Kiyokawa2014}.

The panels (b), (c) and (d) of Fig.\ref{fig:FeLTE} show the CSDs of Fe plasma in condition of $\rho$=0.008 g/cm$^{3}$ and $T$=25, 50, 100 eV respectively. In general, our calculated CSDs agree with those of AAM. At 25eV the small difference between the most populated charge states Fe$^{7+}\& $ Fe$^{8+}$  should result from the different atomic data such at the ionization energies and oscillator strengths. At 100eV the bigger difference of the fractions of Fe$^{16+}\&$ Fe$^{17+}$ may be from the critical situation that the L shell is opening, which is sensitive to the ionization and recombination rate coefficients of the near full occupation L shell. Our high fraction of Fe$^{17+}$ means the L shell is opened earlier than AAM.

\begin{figure}
\includegraphics[scale=0.35]{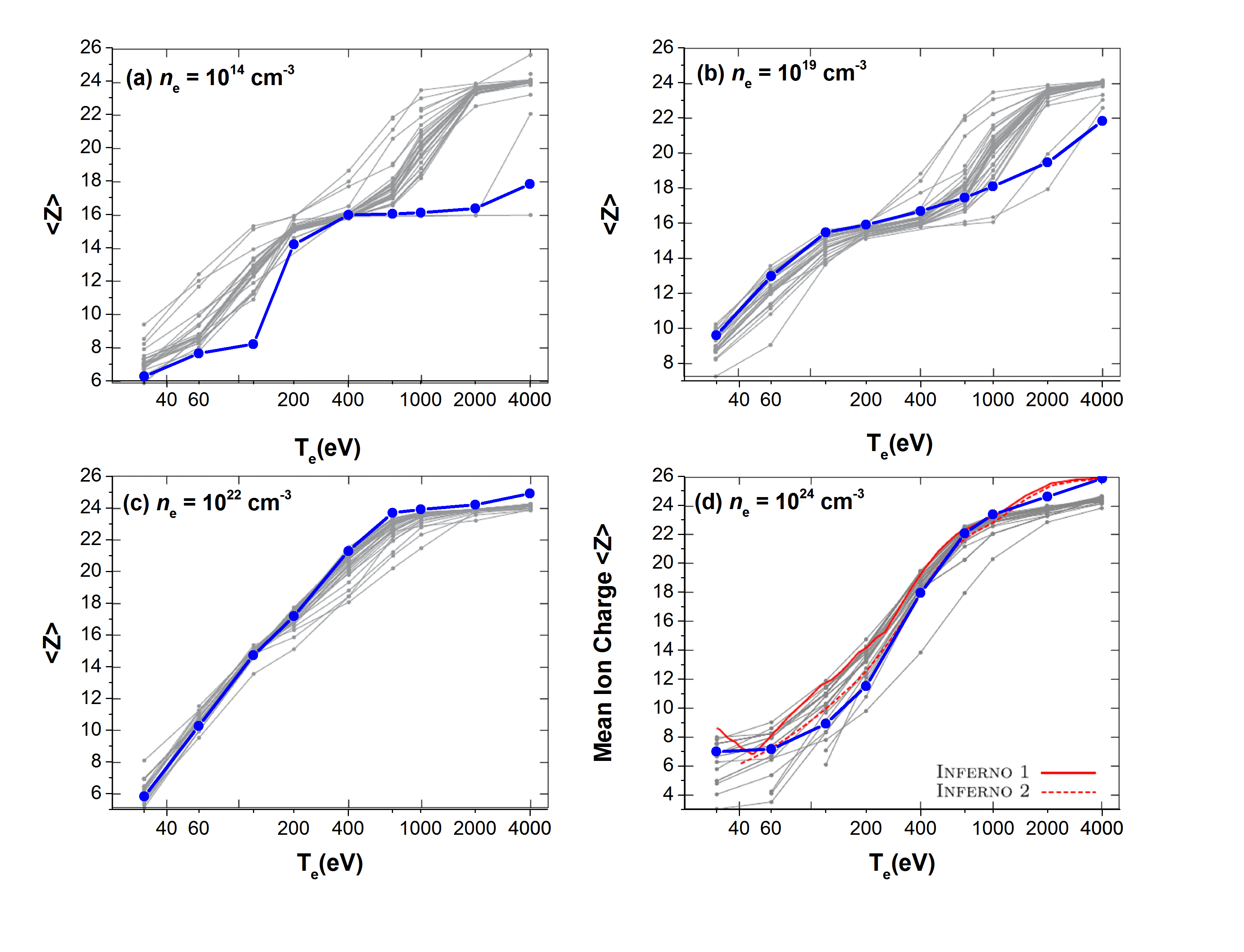}
\caption{\label{fig:FeNLTE}(Color online) The comparison of the mean ionization of Fe Plasma as a function of free-electron temperature $T_{e}$ in NLTE condition between MAICRM results (blue dot) and the results of 14 codes and models submitted in the 3th NLTE code comparison workshop\cite{NLTE3-2006}. }
\end{figure}

Fig.\ref{fig:FeNLTE} shows the mean ion charge $\langle Z\rangle$ of Fe plasma in NLTE condition of $n_{e}=10^{14}, 10^{19}, 10^{22}, 10^{24}$ cm$^{-3}$ respectively. Our results are compared with the data from the various NLTE codes submitted in the 9th NLTE code comparison workshop\cite{NLTE9-2017}, which include a few categories, including AAM codes, configuration and/or Superconfiguration codes, detailed level accounting codes, and hybrid codes. Panel (a) shows at the very low density $n_{e}=10^{14}$ cm$^{-3}$ our calculated mean ionization is generally lower than the others, while at higher electron densities $n_{e}=10^{19}, 10^{22}, 10^{24}$ cm$^{-3}$ our calculated $\langle Z\rangle$ agree with the other results as shown in panels (b),(c) and (d). In MAICRM, the mean occupation numbers are mainly determined by the competition between the excitation and de-excitation channels, which assumes the change of $\mathbf{\Omega}^{\theta}_{i}$ due to the ionization and recombination is a small perturbation and negligible. It is known that without the radiation field, the excitation channels are dominated by the collisional excitation (CE) and the de-excitation channels are dominated by spontaneous emission (SE) as the electron density is low. At very low electron density, the CE rates are much smaller than the SE rates, so the electrons are accumulated in the lowest orbitals and the excited orbitals are nearly empty. Because of the near empty occupation of the excited orbitals, the recombination processes, especially the resonant EC process, are prominent and their contribution to $\mathbf{\Omega}^{\theta}_{i}$ for the excited orbitals is not negligible. If considering the contribution of the recombination, $\mathbf{\Omega}^{\theta}_{i}$ of the excited orbitals should increase a small but non-negligible value comparing with their tiny absolute values, which will result in an obvious increase of AI rates. So the mean ionization of MAICRM at low density is lower than other models. How to treat the very low density situation reasonably deserves further study and will be reported elsewhere. At the mid- and high-densities, the collisional excitation/de-excitation channels become important and much excited states are produced so $\mathbf{\Omega}^{\theta}_{i}$ is a good estimation for the most populated configurations in one charge state, which is illustrated by the good agreement between the MAICRM calculations and the other code results shown in panels (b),(c) and (d).

\begin{figure}
\includegraphics[scale=0.35]{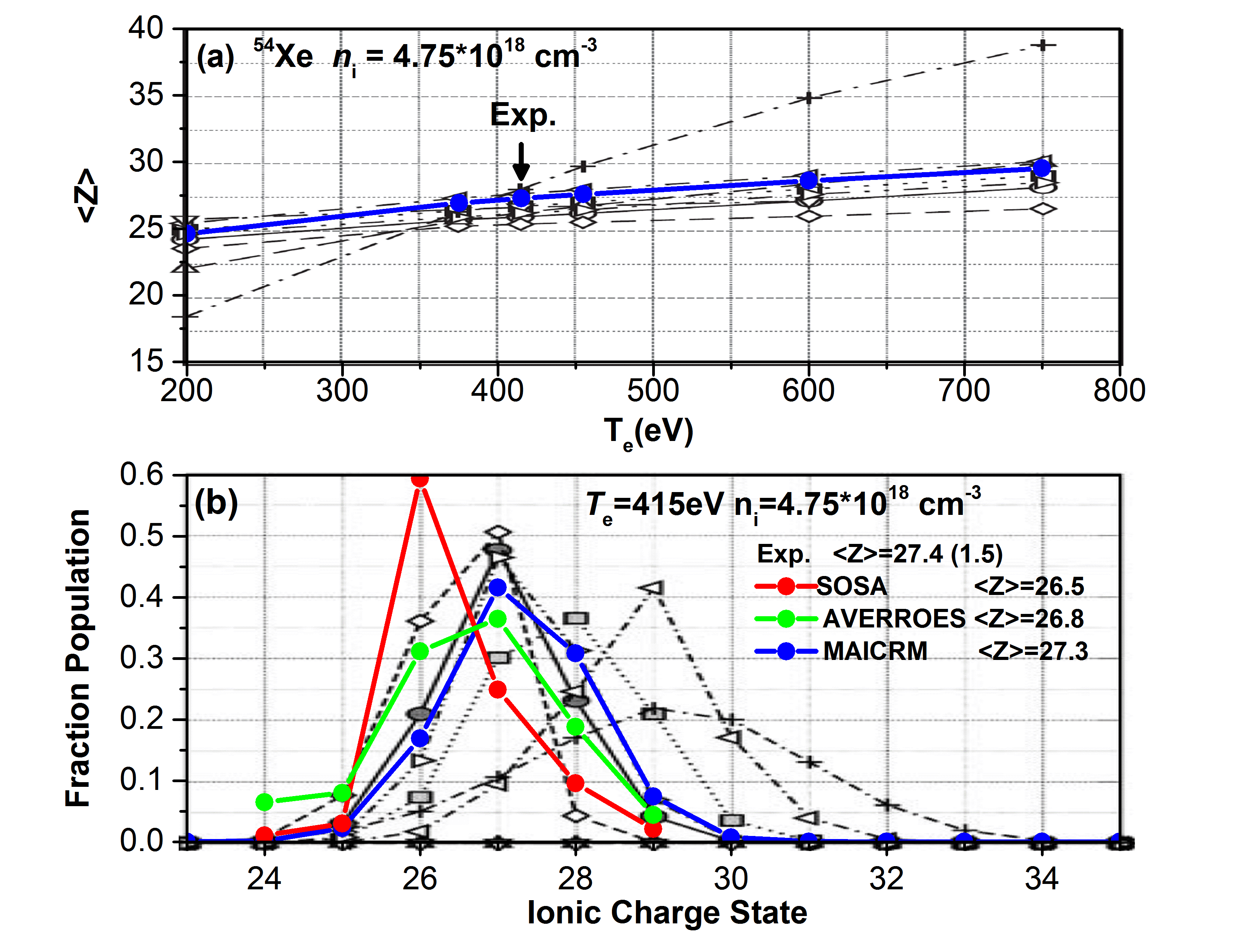}
\caption{\label{fig:Xe}(Color online) (a) The comparison of the mean ionization of Xe plasma as a function of free-electron temperature $T_{e}$ in NLTE condition between MAICRM results (blue dot) and the results of 14 codes and models submitted in the 3th NLTE code comparison workshop \cite{NLTE3-2006} from superconfigurations to detailed level accounting. (b) Charge state distribution of Xe plasmas at $T_{e}=415eV$ and $n_{i}=4.75\times10^{18}$ cm$^{-3}$. The experimental value is $\langle Z\rangle =27.4\pm 1.5$\cite{Popovics2002} and the data in parenthesis are the experimental uncertainties; (red dot): the ion-by-ion spin-orbit-split (SOSA) approach\cite{SOSA1991}; (green dot): the AVERROES calculations based on the superconfiguration and STA concepts\cite{Peyrusse2000}. The other data (gray) are the results of 14 codes and models submitted in the 3th NLTE code comparison workshop \cite{NLTE3-2006}.}
\end{figure}

Fig.\ref{fig:Xe} shows the mean ionization of Xe NLTE plasma at $n_{i}=4.75\times10^{18}$ cm$^{-3}$ as a function of free-electron temperature $T_{e}$ and the CSD at $T_{e}$=415eV. Panel (a) shows the MAICRM mean ionizations are in the spread range of 14 codes submitted in the 3th NLTE code comparison workshop\cite{NLTE3-2006} from superconfigurations to detailed level accounting. Panel (b) shows the CSD of Xe plasmas at $T_{e}=415eV$ and $n_{i}=4.75\times10^{18}$ cm$^{-3}$.  $\langle Z\rangle =27.3$ from the MAICRM calculation agrees better with the experimental $\langle Z\rangle =27.4\pm 1.5$\cite{Popovics2002} than the ion-by-ion spin-orbit-split (SOSA) approach\cite{SOSA1991} and the AVERROES calculations based on the superconfiguration and STA concepts\cite{Peyrusse2000}. The CSD of MAICRM is in the spread range of the results gathered in the 3th NLTE code comparison workshop\cite{NLTE3-2006}.

\begin{figure}
\includegraphics[scale=0.35]{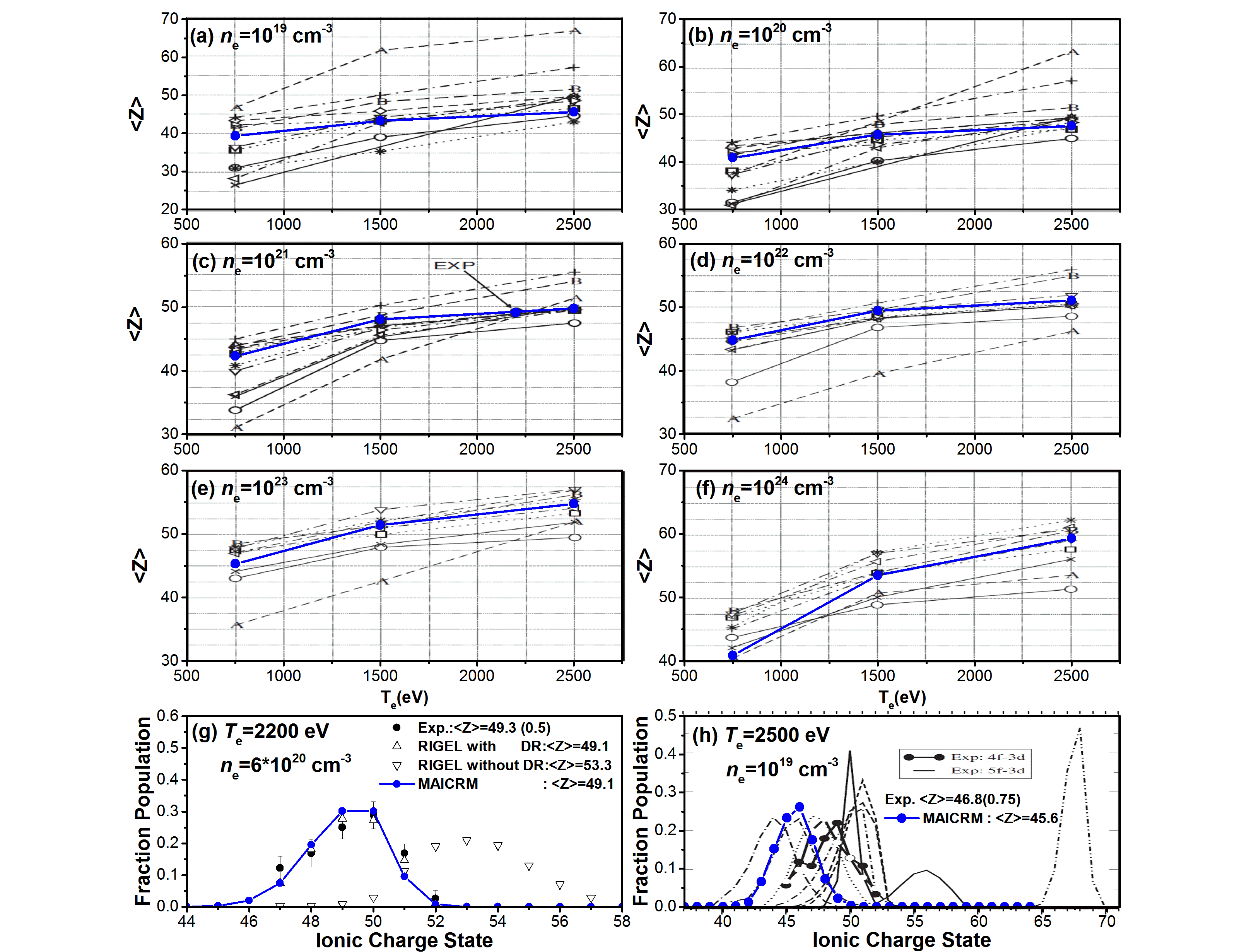}
\label{fig:Au1}\caption{(Color online) The comparisons of the mean ion charge $\langle Z\rangle$ and CSDs of Au plasma between MAICRM and the data from the 9th NLTE code comparison workshop\cite{NLTE3-2006} and Ref.\cite{Foord2000}. Note that in panels (g) and (h) the data in parenthesis of the experimental $\langle Z\rangle$ are the experimental uncertainties. }
\end{figure}

In Fig.\ref{fig:Au1} the panels (a)-(f)  show the mean ionizations of Au plasma at $n_{e}=10^{19}, 10^{20}, 10^{21}, 10^{22}, 10^{23}, 10^{24}$ cm$^{-3}$ as a function of free-electron temperature $T_{e}$. All the mean ionization curves of MAICRM are in the spread range of the submission data from 14 different codes in the 3th NLTE code comparison workshop\cite{NLTE3-2006}. (g) and (h) panels show the CSDs of Au plasma at $T_{e}$=2.2keV $n_{e}=6\times 10^{20}$ cm$^{-3}$and $T_{e}$=2.5keV $n_{e}=10^{19}$ cm$^{-3}$ respectively. As shown in panel (g) the CSD of MAICRM agree well with experimental data and the theoretical values of RIGEL including the two-electron AI/EC processes\cite{Foord2000}, where RIGEL is a superconfiguration-based collisional-radiative code constructed using hydrogenic supershells. Panel (h) shows the mean ionization of MAICRM is lower than the experimental value while in the spread rang of the theoretical data in in the 3th NLTE workshop.

\begin{figure}
\includegraphics[scale=0.35]{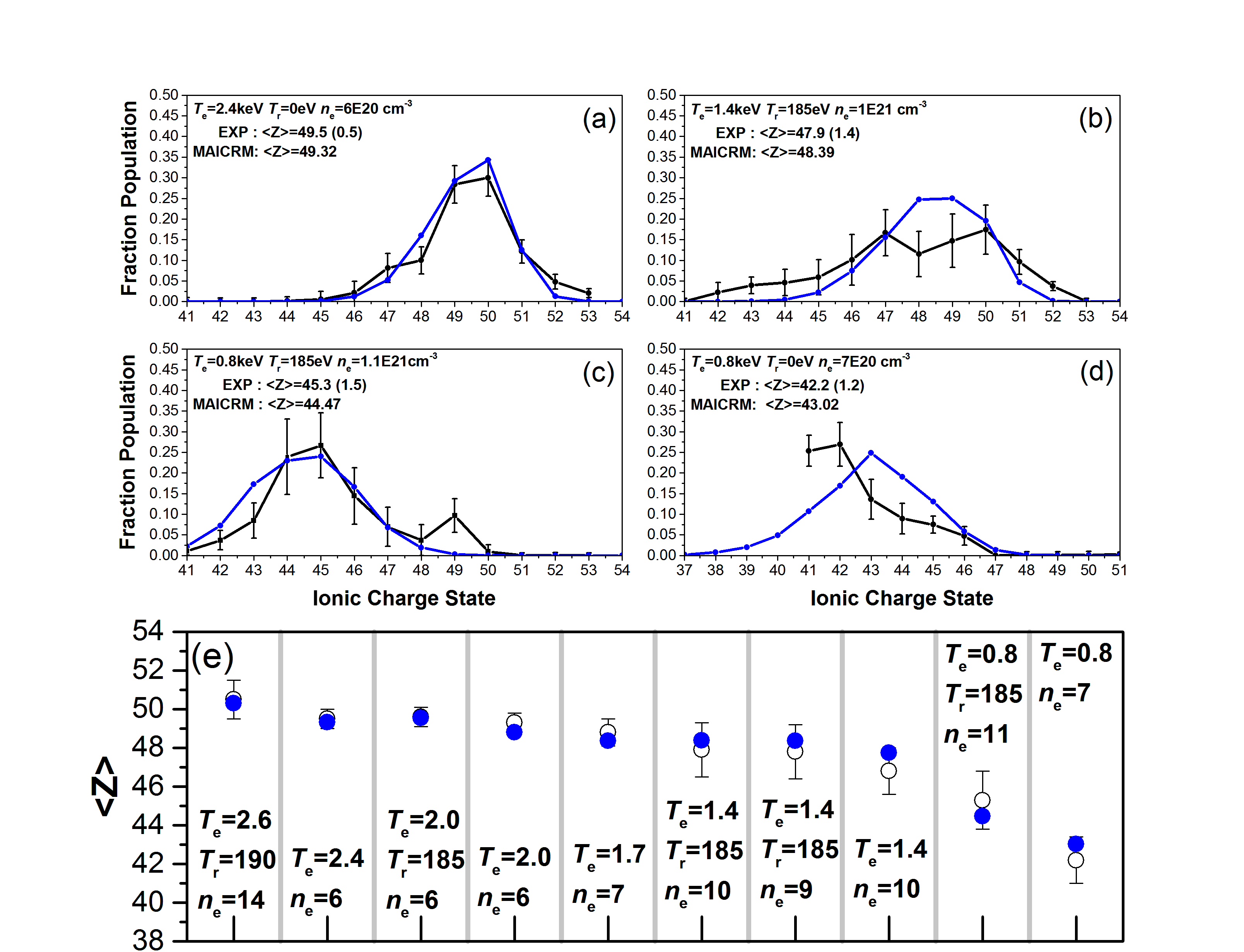}
\label{fig:Au2}\caption{(Color online) The comparisons of the mean ion charge $\langle Z\rangle$ and CSDs of Au plasma between MAICRM and the experimental results\cite{HeeterPRL2007}. Note that in panels (a), (b), (c) and (d) the data in parenthesis of the experimental $\langle Z\rangle$ are the experimental uncertainties and in panel (e) the units of electron temperature $T_{e}$, radiation temperature $T_{r}$ and electron density $n_{e}$ are keV, eV and $10^{20}$ cm$^{-3}$ respective.}
\end{figure}

Fig.\ref{fig:Au2} shows the comparisons of the MAICRM mean ionization and CSDs of Au plasma with the recent experimental values\cite{HeeterPRL2007}. In panels (a)-(c) the MAICRM CSDs generally agree with the experimental values. In panel (d) the difference between the MAICRM results and the experimental data is due to a population fraction of 0.15 for the ions with Z$\leqslant$ 40 not modeled in the experimental fitting procedure. Panel (e) shows the comparisons of the mean ionizations at 10 different conditions between the MAICRM calculations and the experimental values\cite{HeeterPRL2007}. As a whole the MAICRM results agree with experimental values within the experimental uncertainties, which illustrates the credibility of MAICRM for high-Z hot dense plasma.

\section{Conclusion}
A general model MAICRM for the simulation of hot dense plasmas is proposed. In this model, all the energy levels at the same charge stage are averaged and represented by an average ion $\mathbf{\Lambda}^{\theta}$ and (Z+1) average ions are used to describe the plasma. The average occupation numbers $\mathbf{\Omega}^{\theta}_{i}$ are determined by the competition between the excitation and de-excitation processes in the average ion $\mathbf{\Lambda^{\theta}}$ and the populations $\mathbf{P}_{\theta}$ for each average ion are determined by the ionization and recombination processes between the average ions. After iteratively solving the two sets of rate equations a set of converged $\{ \mathbf{\Omega}^{\theta}_{i},\mathbf{P}_{\theta},i\in(1, i^{\theta}_{max}),\theta\in(0,Z)\}$ are obtained. The calculation results of MAICRM at mid- and high-density conditions agree with the other more detailed theoretical results and the experimental results. For the very low density condition the mean ionization of MAICRM is lower than the other theoretical results since the recombination rate coefficients are larger than the excitation and de-excitation rates, which situation deserves further study. Considering the calculation time, since the $(N_{CS}\times N_{C})^{\textsc{MAICRM}} = (Z+1)$, and $(N_{CS}\times N_{C})^{\textsc{AAM}}=1$, MAICRM is slower than AAM, while MAICRM is faster than the other detailed models such as DCA or SCA because of $(N_{CS}\times N_{C/SC})^{\textsc{DCA/SCA}}=(Z+1)\times N_{C/SC}$. On the other hand MAICRM can treat more detailed transitions and ionization balances than AAM. Therefore, MAICRM has the advantage to be combined into 1D, 2D and 3D radiative hydrodynamic calculations of ICF applications with more detailed treatments than AAM and shorter time-consuming than DCA/SCA models.

\begin{acknowledgments}
This work is partly supported by the National Key R\&D Program of China Under Grant No. 2017YFA0402300.
\end{acknowledgments}

\nocite{*}

\begin{thebibliography}{9}
%
\bibitem{Lindl1995}
J. D. Lindl, Phys. Plasmas \textbf{2}, 3933 (1995).
%
\bibitem{Whitney2001}
K. G. Whitney \emph{et al.}, Rev. Sci. Instrum. \textbf{8}, 3708 (2001).
%
\bibitem{NLTE9-2017}
R. Piron, F. Gilleron, Y. Aglitskiy, H.-K. Chung, C. J. Fontes, S. B. Hansen, O. Marchuk, H. A. Scott, E. Stambulchik, Yu. Ralchenko, High Energy Density Phys. \textbf{23}, 38 (2017).
%
\bibitem{UTA1988}
C. Bauche-Arnoult, J. Bauche, and M. Klapisch, Adv. At. Mol. Phys. \textbf{23}, 131 (1988).
%
\bibitem{STA1989}
A. Bar-Shalom, J. Oreg, W. H. Goldstein, D. Shvarts, and A. Zigler, Phys. Rev. A \textbf{40}, 3183 (1989).
%
\bibitem{Ralchenko2001}
Yu. Ralchenko, J. Quant. Spectrosc. Radiat. Transf. \textbf{71}, 609-621 (2001).
%
\bibitem{SCRAM2007}
S. S. B. Hansen, J. Bauche, C. Bauche-Arnoult, and M. F. Gu, High Energy Density Phys. \textbf{3}, 109 (2007).
%
\bibitem{DCA1990}
B. Wilson, J. Albritton, and D. Liberman, "Detailed configuration account opacity calculations using a Monte Carlo technique", LLNL Report No. UCRL-JC-014252 (1990).
%
\bibitem{JONES2017}
O. S. Jones, L. J. Suter, H. A. Scott, M. A. Barrios, W. A. Farmer, S. B. Hansen, D. A. Liedahl, C. W. Mauche, A. S. Moore, M. D. Rosen, J. D. Salmonson, D. J. Strozzi, C. A. Thomas, and D. P. Turnbull, Physics of Plasmas \textbf{24}, 056312 (2017).
%
\bibitem{TXM1995}
X. M. Tong, Y. Zou and J. M. Li, Chin. Phys. Lett. \textbf{12}, 351 (1995).
%
\bibitem{Florido2009}
F. Florido \emph{et al.}, Phys. Rev. E \textbf{80}, 056402 (2009).
%
\bibitem{Stewart1966}
J. C. Stewart and K. D. Pyatt, Astrophys. J. \textbf{144}, 1203 (1966).
%
\bibitem{Faussurier1997}
G. Faussurier, C. Blancard, and B. A. Decoster, Phys. Rev. E. \textbf{56}, 3474 (1997).
%
\bibitem{Kiyokawa2014}
Shuji Kiyokawa, High Energy Density Phys. \textbf{13}, 40 (2014).
%
\bibitem{Popovics2002}
C. Chenais-Popovics, J. C. Gauthier, J. C. Gary, O. Peyrusse, M. Rabec-Le Gloahec \emph{et al.},  Phys. Rev. E \textbf{65}, 046418 (2002).
%
\bibitem{SOSA1991}
J. Bauche, C. Bauche-Arnoult, and M. Klapisch, J. Phys. B \textbf{24}, 1 (1991).
%
\bibitem{Peyrusse2000}
O. Peyrusse, J. Phys. B \textbf{33}, 4303 (2000).
%
\bibitem{NLTE3-2006}
C. Bowen, R. W. Lee, and Yu. Ralchenko, JQSRT \textbf{99}, 102-119 (2006).
%
\bibitem{Foord2000}
M. E. Foord, S. H. Glenzer, R. S. Thoe, K. L Wong, K. B. Fournier, B. G. Wilson, and P. T. Springer, Phys. Rev. Lett. \textbf{85}, 992 (2000).
%
\bibitem{HeeterPRL2007}
R. F. Heeter, S. B. Hansen, K. B. Fournier, M. E. Foord, D. H. Froula, A. J. Mackinnon, M. J. May, M. B. Schneider, and B. K. F. Young, Phys. Rev. Lett. \textbf{99}, 195001 (2007).
%
%
%
%
%
%
%
%
%
%
%
%
%
%
%
\end{thebibliography}

\end{document}